\documentclass{vietnam}

\def\be{\begin{equation}}
\def\ee{\end{equation}}
\def\bea{\begin{eqnarray}}
\def\eea{\end{eqnarray}}

\newcommand{\Photo}{\includegraphics[height=35mm]{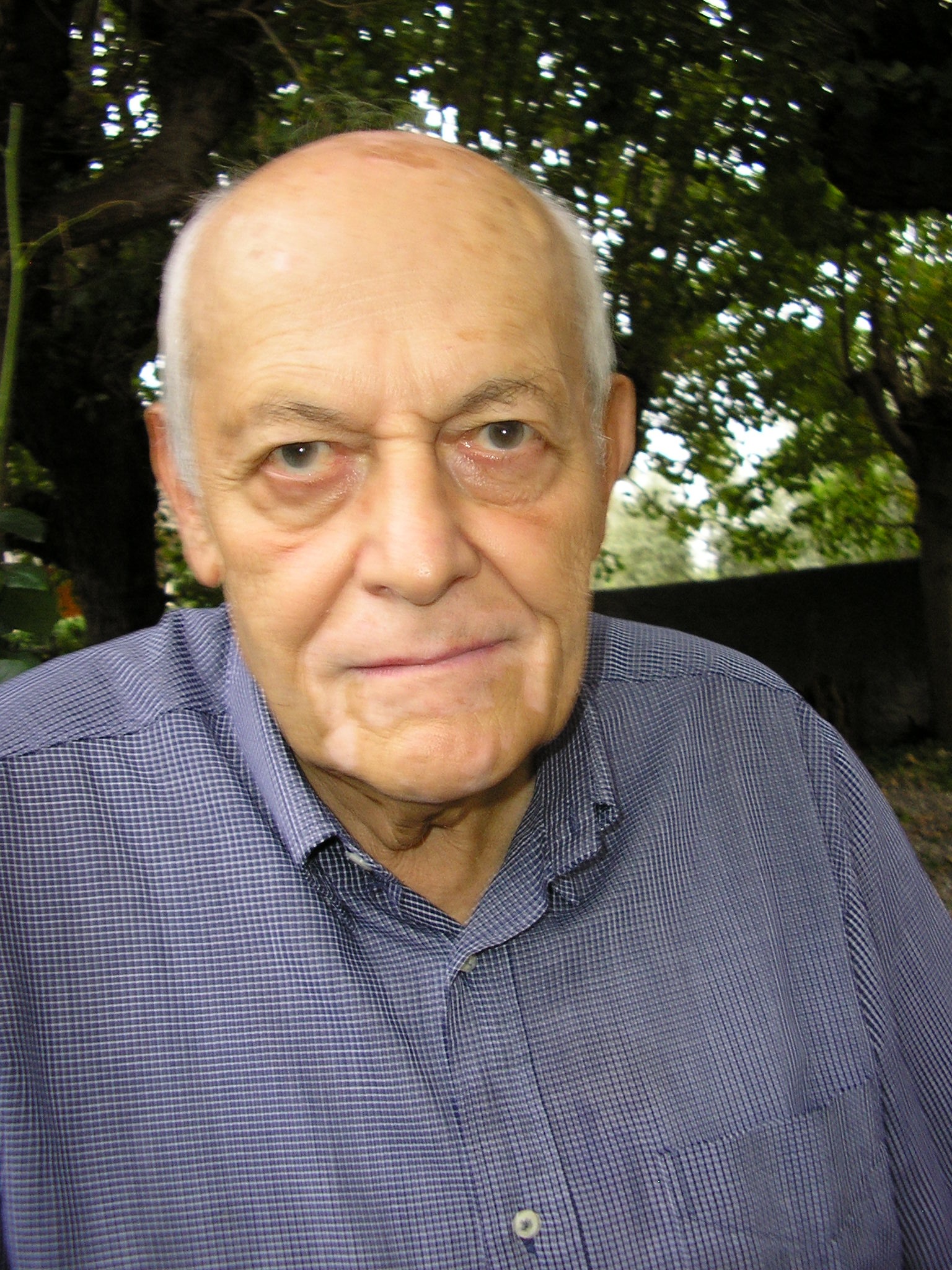}}

\begin{document}
\vspace*{4cm}
\title{Understanding Color Confinement}

\author{ A. Di Giacomo }

\address{Dipartimento di Fisica Universita' di Pisa and I.N.F.N. Sezione di Pisa\\ 3 Largo B. Pontecorvo
56127 Pisa ITALY}

\maketitle\abstracts{Some aspects are discussed of the mechanism of color confinement in $QCD$ by
condensation of magnetic monopoles in the vacuum.
}

\section{Introduction}
 $QCD$ is an $SU(3)$ gauge theory coupled to quarks in the fundamental representation which describes the strong interaction sector of the Standard Model. The theory has a physical scale $\Lambda \approx 250$ Mev: at distances smaller than $\frac{1}{\Lambda}$ the perturbative expansion works, and the theory behaves as any other field theory; at larger distances the quanta of the fundamental fields, quarks and gluons, do not propagate  as a free particles (Colour  Confinement). 
 
  In nature the ratio of the abundance $n_q$ of quarks with respect to that of protons  $n_p$ has an upper limit
 $\frac{n_q}{n_p} \le 10^{-27}$ to be compared to the expectation in the Standard Cosmological Model $\frac{n_q}{n_p} \approx 10^{-12}$; the ratio of the inclusive cross section $\sigma_q$ for production of quarks in hadronic reactions divided by the total cross section $\sigma_T$  has upper limit  $\frac{\sigma_q}{\sigma_T} \leq 10^{-15}$ and should be  of the order of unity. The only natural explanation of such a strong  inhibition factor ( $10^{-15}$) is that $n_q$ and $\sigma_q$ are strictly zero due to some symmetry.
 
 If this is true the transition to quark-gluon plasma is a change of symmetry, hence a real phase transition and not a cross-over. The transition is observed in  lattice simulations \cite{phil} \cite{ddp}.
 and possibly  in heavy ion collisions at RIHC and CERN.   
 
 The symmetry responsible for confinement cannot be a subgroup of the gauge group, since gauge symmetry is not broken neither below nor above the transition, except in the pure gauge theory ( no quarks) which is blind to the centre $C$. Lattice simulations show that the symmetry $C$  is  spontaneously broken below some temperature $T_c$ ( confined phase), and is restored  above it.  The order parameter is the Polyakov loop $ \langle L \rangle =\frac{1}{V} \int_V d^3x \langle  Tr L(\vec x,t) \rangle$, where
 \begin{equation}
  L(\vec x, t) = P \exp(  i \int^{t+\frac{1}{T}}_{t} dx_4 gA_{4}(\vec x, x_4) ) \ \label{pol}
 \end{equation}
 is the parallel transport along the time axis, which at finite temperature T is a closed loop due to the  periodic boundary conditions. It transforms covariantly, and commutes with the centre of the group, and therefore is zero in the spontaneously broken phase. On the other hand 
 it can be shown that  $\langle L \rangle = \exp( - \frac{F_q}{T} )$ with $F_q$ the free energy of a static quark, so that when $\langle L \rangle =0$ $F_q = \infty$ (confinement). Numerical simulations show indeed that, for pure gauge theories [ $SU(2)$ , $SU(3)$], $T_c \approx 200 Mev $,  $\langle L \rangle = 0$ for $T \le T_c$ and $\langle L \rangle \neq 0$ for $T\ge T_c$.
  However $C$ can not be the symmetry producing confinement in nature, since it is not a symmetry in presence of quarks.  Nor the symmetry can be chiral symmetry, which is not defined in the quenched case and is broken by the masses of the quarks.
  
  The only way to have an extra symmetry in $QCD$ rests on the degrees of freedom on the boundary of the physical space (dual variables). This is what happens in many models of statistical mechanics, like the $2d$ Ising model \cite{Kadanoff}, the $3d$  $ X-Y$ model \cite{Marino}, the lattice $U(1)$ gauge theory\cite{Froel} \cite{dp}.  The nature of the excitations is determined by the dimensions of space. They are kinks in $d=1 +1$ (Ising), vortices in $d= 2+1$, monopoles
  for $d= 3+1$. Monopoles do exist in gauge theories. \cite{'tH} \cite{pol}.
 The idea is physically attractive, since monopoles could condense in the vacuum and make of it a dual superconductor confining chromo-electric charges (quarks) \cite{'tH2} \cite{man}.
 
 \section{Monopoles}

Monopoles as static classical solutions (solitons) do exist in the $SU(2)$ Higgs model.
\begin{equation}
L = -\frac{1}{4} \vec G_{\mu \nu} \vec G_{\mu \nu} + \frac{1}{2}(D_{\mu} \vec \Phi)^{\dagger}(D_{\mu}\vec \Phi) -\frac{\lambda}{4}[\vec \Phi \vec \Phi - \mu^2] ^2 \label{L}
\end{equation}
The equations for static solutions are: \hspace{.9cm}
 $D_{j} G_{j i} = g \vec \Phi \wedge D_{i} \vec \Phi - g \vec A_0 \wedge D_{i} \vec A_0$,   
 
 \hspace{.9cm} $D_{i}D_{i} \vec A_0 =g \vec \Phi \wedge D_0 \vec \Phi$,  \hspace{2.0cm}    $D_0D_0 \vec \Phi - D_i D_i \vec \Phi + \lambda ( \vec \Phi^2 - \mu^2) \vec \Phi =0$.
 
\vspace{.3cm}
If $A_0$ can be gauged away the equations have a monopole solution \cite{'tH} \cite{pol} and read :

\vspace{.3cm}
\hspace{1.0cm}$D_{j} G_{j i} = g \vec \Phi \wedge D_{i} \vec \Phi$, \hspace{2.5cm}$ D_i D_i \vec \Phi + \lambda ( \vec \Phi^2 - \mu^2) \vec \Phi =0$
\vspace{.3cm}

If there is no Higgs field, as is the case in $QCD$, the equations become:

\vspace{.3cm}
\hspace{.8cm} $D_{j} G_{j i} = - g \vec A_0 \wedge D_{i} \vec A_0$, \hspace{1.8cm}$D_{i}D_{i} \vec A_0 = 0 $
\vspace{.3cm}

showing that  $A_4 = i A_0$ can act as effective Higgs field and allow stable monopole solutions. As appears from the equations above the solutions correspond to the case $\lambda=0$ \cite{B} \cite{PS}.
The explicit form of the solution is, in the hedgehog gauge,

\vspace{.3cm}

  \hspace{.9cm}$ A^a_i = \epsilon_{ain}\frac{\hat r^n}{gr}(1- K(\xi) ) $ \hspace{1.80cm} $ A_4^a = \frac{\hat r^a}{gr} J(\xi) $\hspace{1.3cm}  $\xi = g \mu r$
  
  \vspace{.3cm}
  
   \hspace{.9cm}$ K(\xi) = \frac {\xi}{\sinh \xi} $\hspace{3,4cm}  $ J(\xi) = \xi \coth \xi -1 $\hspace{0.3cm} $A_4^a( r =\infty) = \mu \hat r^a$
   
   \vspace{.3cm}
   
   The field strengths at large distances are
   
   \vspace{.3cm}
   
   \hspace{.7cm} $B_i^a \approx_{r \to \infty} \hat r^a \frac{\hat r^i}{gr^2}$ \hspace{2.8cm}
$E^a_i \approx_{r \to \infty} i \hat r^a \frac{\hat r^i}{gr^2}$

\vspace{.3cm}
1) The unitary gauge in which the magnetic field is abelian is the one in which $A_4$ is diagonal. To
     have a monopole $A_4$ must be $\neq0$ at large $r$.

2) The scale is fixed by the value of $|A_4|$ at large distances, which is independent of the direction
     since the configuration has finite energy\cite{GPY}. The mass is  $ M= \frac{4 \pi}{g} \mu$.
     
 3) The configuration is called a dyon and considered electrically charged. In fact the electric field is 
      imaginary at large distances, like in instantons. The configuration rather describes a tunnelling 
      between  states with different magnetic charge.  
 \section{Polyakov loops}
 In the gauge $\partial _4 A_4 =0$ , $A_4 (\vec x) = |A_4| \frac{\sigma_3}{2}$ the Polyakov line Eq.(\ref{pol}) reads
 
$
 L(\vec x) = \exp ( \frac{igA_4(\vec x)}{T})$ \hspace{.3cm} 
 and \hspace{.3cm} $\frac{1}{N} Tr [ L(\vec x) ] = \cos( \frac{g |A_4(\vec x)|}{2T})$.
 
 Here we will consider for the sake of simplicity a gauge theory with gauge group $SU(2)$: the extension to the case of generic group is straightforward.
 
 At vanishing temperature $ T \ll \Lambda$ the time extension of the system is much larger than the correlation length $\frac{1}{\Lambda}$. We can divide it e.g. into two parts, both much larger than the correlation length. The Polyakov line $L$ will be the ordered product of the parallel transports  across the two parts  $L = L_1 L_2$ The boundary configuration between the two parts is irrelevant if the extensions are both $\gg \frac{1}{\Lambda}$, so we can take it periodic and $L_1$ , $L_2$ as Polyakov lines, both at vanishing $T$. Since the two regions of times which are support of $L_1$, $L_2$ are uncorrelated, we get then that  $\langle L \rangle \approx \langle L \rangle^2 $, implying that, if $\langle L \rangle$ has not the trivial value $1$, at zero $T$  $\langle L \rangle=0$. This argument extends to the general case what is true in pure gauge theory because of the invariance under the centre $C$ of the group. Indeed, since the Hilbert space is invariant under $C$ , 
 $L = CL$, $\langle L \rangle = \exp( i \pi ) \times \langle L \rangle =0$ and, as for the spectrum, 
  $0= \langle L \rangle =\int _{-1}^{1} d \cos(x) f(cos(x)) $ : 
  
  $f$ is an odd function of $\cos(x)$.
  
 Since at the deconfining transition $T \approx \Lambda$ we can conclude for the spatial average 
 \begin{equation}
 \langle \cos( \frac{g |A_4(\vec x)|}{2\Lambda}) \rangle _V =0 \label{AVL}
 \end{equation}
 
  If we assume that monopoles, as quanta of the dual fields have all the same mass $ M= \frac{4 \pi}{g} \mu$, and hence the same size and $A_4^a( r =\infty)$,  since for a dilute gas of monopoles the average in Eq.(\ref{AVL}) is dominated by large distances and there is no confinement, we get
  \begin{equation}
  \frac{\mu g}{2\Lambda} = 2k \pi
  \end{equation}
  The size of the monopoles is substantially smaller than $\frac{1}{\Lambda}$ and their mass $M \gg \Lambda$.
  To have confinement monopoles should not be dilute, or the average distance between monopoles should be larger but not too much compared to their size. The simplest model which can give an idea of the orders of magnitude is a gas of non interacting monopoles and anti-monopoles, with density $\rho$, and with the additional simplification that the field $A_4(\vec x)$ around one monopole is that corresponding to an isolated monopole up to the distance of the nearest one.
  
   With these simplifying assumptions the average Eq.(\ref{AVL}) can be computed using the explicit solution of Section 2,  and with it the distribution $f(\cos(x))$ of the Polyakov loop\cite{dig}. The result is
  \begin{equation}
  \langle L \rangle = \int dV P(V) \frac{1}{V}\int dV' \cos [\pi( \coth \xi' - \frac{1}{\xi'})] \label{distr}
  \end{equation}
  with $P(V) = 2 \rho \exp(- 2 \rho V) $   the probability that in a sphere of volume $V$ centred on a monopole there is no monopole (and no anti-monopole whence the factor $2$ ). Explicitly
 
 \vspace{.3cm} 
 
 $\langle L \rangle =  3 A \int_{0}^{\infty} d\xi \xi^2 \ (-) Ei( - A \xi^3)  \cos[\pi(\coth \xi - \frac{1}{\xi})] $ .\hspace{.5cm} $ A = \frac{8\pi}{3} \frac{1}{(g \mu)^3}  \rho = \frac{\rho}{3\pi^2\Lambda^3}$
 
  \vspace{.3cm}
 
 is the number of monopoles plus anti-monopoles in a volume of the size of a monopole.
 
 In Eq.(\ref{distr}) the argument of the cosine ranges from $0$ at $\xi =0$ to $\pi$ as $\xi \to \infty$: at short distances the model is credible, at large distances the details of the multi-monopole configuration are important, but the factor $A$ in front of the result comes from the normalization, which is affected by both regions. The distribution in $z= \cos\theta $ , $\theta =\pi [ \coth \xi -\frac{1}{\xi}]$ reads
 $f (z) =\frac{ 3A B \xi^2 Ei(-A \xi^3)}{\pi (\frac{1}{\xi^2} -\frac{1}{\sinh^2 \xi}) \sin [\pi( \coth \xi - \frac{1}{\xi})]}$, it can be fitted to the lattice data, and $A$, $B$ can be determined. The factor $B$ in front is allowed to correct for the normalisation, which can be determined from the region $z\ge 0$, which is more credible. A reasonable agreement is found for $B\approx 2.3$ $A\approx .02$. 
 If monopoles condense their density $\rho$ is related to the $vev$   $<\Phi^2> = \frac{M^2}{\lambda}$, as $\rho = M<\Phi^2> = \frac{M^3}{\lambda}$ with $M = \frac{4 \pi}{g}\mu$ the known mass of the monopole and $\lambda$ the coupling of the quartic term in the effective lagrangean. From the determination  of $A$  the ratio  $\eta = \frac{1}{g^2 \lambda}$ between the magnetic coupling constant $\frac{1}{g}$ and the quartic coupling $\lambda$ can be extracted, getting  $ g \lambda^{\frac{1}{2}} = \frac{2}{3}\frac{(4 \pi)^2}{A g^4}$, indicating a value larger than $\sqrt 2$, i.e. a type II superconductor.
 \section{Discussion}
 Monopole dominance\cite{Suzuki} and an approach based on symmetry \cite{dp} \cite{dlmp,bcdd}, strongly support monopole condensation in the vacuum as mechanism of colour confinement. A revival of this idea recently came from the study of instantons with non-trivial holonomy, named calorons,  \cite{KvB} \cite{Diakonov}, which prove to have monopoles as constituents.
 
 Lattice configurations contain monopoles which propagate on distances of the order of the lattice spacing, which can be considered as fluctuations, and monopoles which propagate on distances of the order of the correlation length and larger, which can be considered as stable, and which can condense. To have stable monopoles a Higgs fleld is needed. In $QCD$ there is no fundamental Higgs field, but the euclidean time component of the gauge field, $A_4$ can act  as an effective Higgs field.
 Stable monopoles result, which have a definite form and a scale which is determined by holonomy, i.e. by the value of $A_4$ at large distances. The Polyakov line Eq.(\ref{pol}) plays then a fundamental role in confinement, even in systems with quarks, where there is no invariance under the centre $C$ of the gauge group, being related to the existence of stable monopoles which have to produce dual superconductivity.
 
 An attempt to describe the monopoles as constituents of calorons forming a gas proves to be unviable in the non-perturbative region, relevant to physics\cite{Diakonov}. We have sketched a much simpler approach treating monopoles as independent particle. The model is better than it looks at first sight, since magnetic charges are shielded by condensation in the confined phase.
 
 A detailed presentation and the extension to finite temperature will be presented in Ref.\cite{dig}.


\section*{References}

\begin{thebibliography}{99}
\bibitem{phil} For a review see e.g. O.~Philipsen, Prog.Theor.Phys.Suppl. 174 (2008) 206
\bibitem{ddp} M.~D'Elia, A.~Di Giacomo,C.~Pica  Phys.Rev. D72 114510 (2005) 
 \bibitem{Kadanoff}L.~P.~Kadanoff, H.~Ceva, Phys.Rev. B3 (1971) 3918
 \bibitem{Marino}T.~Banks, R.~Meyerson, J.B.~Kogut Nucl. Phys {\bf B129} 493 (1977)
 \bibitem{Froel}J.~Froelich, P.~Marchetti,Comm. Math. Phys. {\bf 112}, 343 (1987).
\bibitem{dp} A.~Di~Giacomo, G.~Paffuti, Phys. Rev. D {\bf 56}, 6816 (1997).
\bibitem{'tH} G.~'t~Hooft, Nucl. Phys. B {\bf 79}, 276 (1974).
 \bibitem{pol} A.~M.~Polyakov, JETP Lett.  {\bf 20}, 194 (1974).
 \bibitem{'tH2} G.~'t~Hooft, in \emph{High Energy Physics}, EPS Int.Conf., Palermo 1975, A. Zichichi ed.
 \bibitem{man} S.~Mandelstam, Phys. Rep. {\bf 23C}, 245 (1976).
   \bibitem{B} E.~B.~ Bogomonlyi, Sov.~J.~Nucl.~Phys {\bf 24}, 449 (1976)
  \bibitem{PS} M.~K.~ Prasad, C.~M.~Sommerfeld, Phys. Rev. Lett {\bf 35}, 760 (1973) 
  \bibitem{GPY} D.~J.~Gross,~R.~D.~Pisarski,~L.~G.~Yaffe, Rev.~Mod.~Phys. {\bf 53}, 43 (1981)
  \bibitem{dig} A.~Di Giacomo, in preparation.
  \bibitem{dlmp} A.~Di~Giacomo, B.~Lucini, L.~Montesi, G.~Paffuti. Phys. Rev. D {\bf 61}, 034503 (2000).
\bibitem{bcdd} C.~Bonati, G.~Cossu,M.~D'Elia, A.~Di Giacomo, Phys.Rev.{\bf D85} 065001 (2012)
\bibitem{Suzuki} T.~Suzuki, I.~Yotsuyanagi, Phys. Rev. D {\bf 42}, 4257 (1990).
\bibitem{KvB}T.C.~Kraan, P.~van Baal Phys. Lett. {\b B435}, 399 (1998)
\bibitem{Diakonov}  D.~Diakonov, Acta Phys.Polon. {\bf B39}  3365  (2008)
\end{thebibliography}
\end{document}